\documentstyle[12pt]{article}

\begin{document}

{}
\hfill   {\bf \large IFT/4/99}
\vskip1in
\centerline{\Large {\bf Boost-Invariant Running Couplings in }}
\vskip.1in
\centerline{\Large {\bf Effective Hamiltonians }}
\vskip .1in
\centerline{\small {May, 1999}}
\vskip .3in
\centerline{Stanis{\l}aw D. G{\l}azek}
\vskip .1in
\centerline{Institute of Theoretical Physics, Warsaw University}
\centerline{ul. Ho{\.z}a 69, 00-681 Warsaw}
\vskip.5in

\centerline{\bf Abstract}
\vskip.1in

We apply a boost-invariant similarity renormalization group procedure to
a light-front Hamiltonian of a scalar field $\phi$ of bare mass $\mu$
and interaction term $\sim g \phi^3$ in 6 dimensions using 3rd order
perturbative expansion in powers of the coupling constant $g$.  The
initial Hamiltonian is regulated using momentum dependent factors that
approach 1 when a cutoff parameter $\Delta$ tends to infinity.  The
similarity flow of corresponding effective Hamiltonians is integrated
analytically and two counterterms depending on $\Delta$ are obtained
in the initial Hamiltonian:  a change in $\mu$ and a change of $g$.  In
addition, the interaction vertex requires a $\Delta$-independent
counterterm that contains a boost invariant function of momenta of
particles participating in the interaction.  The resulting effective
Hamiltonians contain a running coupling constant that exhibits
asymptotic freedom.  The evolution of the coupling with changing width
of effective Hamiltonians agrees with results obtained using Feynman
diagrams and dimensional regularization when one identifies the
renormalization scale with the width.  The effective light-front
Schr\"odinger equation is equally valid in a whole class of moving
frames of reference including the infinite momentum frame.  Therefore,
the calculation described here provides an interesting pattern one can
attempt to follow in the case of Hamiltonians applicable in particle
physics.


\newpage

{\bf 1. INTRODUCTION}
\vskip.1in

Similarity renormalization group procedure for Hamiltonians is a method
suggested for seeking solutions to QCD and other quantum field theories
{\it en bloc} including bound states, by calculating manageable
effective Hamiltonians and solving Schr\"odinger equations with them in
a series of successive approximations of increasing accuracy.  This
article describes an elementary application of the similarity method in
case of scalar fields, showing details of a calculation of one
boost-invariant running coupling constant in effective Hamiltonians in
third order perturbation theory.  We briefly review the method, describe
the simple example and present conclusions.

The method originates in the notion of renormalization group as
discussed by Wilson \cite{Wilson1} \cite{Wilson2} and uses the idea of
similarity renormalization group procedure for Hamiltonians
\cite{GlazekWilson1} \cite{GlazekWilson2}.  Similarity enables us to
avoid small energy denominators in perturbative evaluation of effective
Hamiltonians.  The evaluation includes finding counterterms and defining
renormalized dynamics.  Small energy denominators could lead to large
errors in the counterterms and in calculation of effective Hamiltonians.
Thus, if not avoided through similarity, the small denominators would
prevent precise theoretical predictions based on the effective
Schr\"odinger equations.

The similarity procedure was invented to sort out complexities of the
light-front form of Hamiltonian dynamics.  This form was distinguished a
long time ago by Dirac \cite{Dirac} and more recently became a natural
candidate for description of hadronic constituents in hard scattering
processes \cite{BrodskyLepage} as well as in spectroscopy
\cite{Wilsonetal} using QCD.  A recent review article by Brodsky, Pauli
and Pinsky \cite{BrodskyPauliPinsky} provides a description of
theoretical advances made in light-front formulation of various theories
mainly before invention or independently of the similarity procedure.
Reviews by Burkardt, Harindranath and Perry
\cite{BurkardtHarindranathPerry} help in understanding the scope of
current approaches.  Recent research in the direction of renormalization
of Hamiltonians can be traced through Ref.  \cite{RofH}.  The present
article is focused on similarity in light-front dynamics.

Initial studies of quarkonium bound states, which are related to the
similarity program described in \cite{Wilsonetal} have been performed by
Brisudov\'a and Perry, and Brisudov\'a, Perry and Wilson
\cite{BrisudovaPerryWilson} following the key observation by Perry
\cite{Perryconfinement} that second order effective Hamiltonian of QCD
contains a confining term, which may remain uncanceled in the effective
Schr\"odinger equation dynamics.  Higher than second order calculations
are needed for verifying this hypothesis.

Since the formal front form of Hamiltonian dynamics is invariant under
boosts one hopes it can provide a link between the structure of hadrons
at rest and in the parton model.  That such unifying picture is hard to
achieve in standard dynamical schemes is best illustrated by the fact
that despite extensive progress lattice gauge theory does not easily
yield desired quark and gluon bound state wave functions.
\cite{Lattice} The light-front approach is still far from achieving this
goal, too.  The present article shows the essence of boost invariance in
similarity \cite{GlazekAPP} but the example we describe here is limited
to scalar particles.  The dynamics of scalars does not involve genuine
small-$x$ singularities ($x$ denotes a longitudinal momentum fraction
carried by a particle in the infinite momentum frame) that appear in
gauge theories and the present work does not describe known particles.

The present paper also does not cover the step of solving the effective
Schr\"odinger equation.  It was shown in an asymptotically free matrix
model \cite{GlazekWilson3} that one can achieve 10 \% accuracy in
calculating bound state properties using second order effective
Hamiltonians but it is not known yet if the same accuracy can be reached
in any quantum field theory of interest using the boost-invariant
similarity approach discussed here.

Before we proceed to our example, we first disclose the mechanism of
preserving boost invariance in the similarity approach.  The following
discussion points out relevant features of the procedure using other
methods for comparison.

The similarity renormalization group procedure leads to effective
Hamiltonians $H_\lambda = H_{0\lambda} + H_{I\lambda}$, whose matrix
elements between any two eigenstates of $H_{0\lambda}$ vanish when the
eigenvalues of $H_{0\lambda}$ for these states differ by more than the
"width" $\lambda$.  The word "width" is natural because the effective
Hamiltonian matrix can be viewed as a band of non-zero matrix elements
along the diagonal and the width of the band depends on $\lambda$.
Another reason for the word "width" is that the band structure is
ensured by similarity form factors in interaction vertices.  The form
factors are functions of energy transfers, they are peaked around zero
and their half-width depends on $\lambda$.  The notion of the
Hamiltonian width is key to our method of preserving boost invariance.

We shall take advantage of Wegner's equation \cite{Wegner} to indicate
where the boost invariance can enter the similarity procedure.  Wegner
invented a flow equation for diagonalization of Hamiltonian matrices in
solid state physics \cite{Wegner} \cite{Mielke} that is beautifully
simple and can be adapted to the similarity renormalization scheme
\cite{Australia} \cite{GlazekWilson3}.  Wegner's equation for
Hamiltonian matrices can be written as

$$ {d\over ds} \, \, H_\lambda \quad = \quad - \quad [\, [H_\lambda,
H_{0\lambda}], \, H_\lambda] \quad , \eqno(1.1)$$

\noindent where $s = \lambda^{-2}$.  Initial condition should be
provided at $s = 0$, corresponding to $\lambda = \infty$, and the
initial Hamiltonian is denoted by $H_\infty$.  Wegner's commutator
$[H_\lambda, H_{0\lambda}]$ generates the similarity transformation.
One discovers a gaussian similarity factor by solving Eq.  (1.1) for the
$H_{I\lambda}$ matrix elements between eigenstates of matrix $H_{0\lambda}$
keeping on the right-hand side only those terms that are linear in
$H_{I\lambda}$ and neglecting higher order terms.

Wegner's equation preserves necessary cluster decomposition properties
\cite{Weinberg}.  The commutator structure of Wegner's equation implies
that no disconnected interactions are generated by the transformation.
Wegner's transformation depends only on differences of energies and
spectator energies always drop out from the differences.  However,
Wegner's generator is not boost invariant.

Seeking a boost invariant approach, one can apply Eq.  (1.1) to
light-front Hamiltonian matrices.  The matrices are of the form $H =
(P^{\perp\,2} + M^2_\lambda)/P^+$, where the mass matrix contains
interactions, $M^2_\lambda = M^2_{0\lambda} + M^2_{I\lambda}$, and $P$
denotes total momenta of states with which one evaluates the matrix
elements.  Since the masses do not change the momenta, one can rewrite
Eq.  (1.1) in terms of the mass matrix elements as follows

$$ P^{+\,2}\,\,{d \over d s}\,\, M^2_\lambda \quad = \quad - \quad
[\,[M^2_\lambda, M^2_{0\lambda}],\,M^2_\lambda] \quad . \eqno(1.2)$$

\noindent We see that a rescaling of the flow parameter $s$ with a
momentum eigenvalue gives a flow equation for the mass squared matrix
elements.  The latter should be independent of the eigenvalues $P^+$ and
$P^\perp$.  This feature of masses seems to suggest a boost invariant
renormalization procedure based on replacing $P^{+\,2}/s$ by a new
$P^+$-independent flow parameter $\Lambda^4$.  But such substitution
breaks connection with Wegner's equation for Hamiltonian matrices and
leads to violation of cluster decomposition properties, since the mass
is not an additive quantity.  Namely, the mass depends on the relative
motion of particles and effective interactions become dependent on
spectators momentum.

Despite this drawback, Allen and Perry \cite{AllenPerry} succeeded in
demonstration that one can define and calculate a running coupling
constant using Eq.  (1.2) for $M^2$ matrix elements replacing
$P^{+\,2}/s$ with $\Lambda^4$ in massless $\phi^3$ theory in 6
dimensions.  The authors demand that the flow from some $\Lambda$ to a
fraction of $\Lambda$ reproduces the same matrix elements structure in
which only some parameters change.  This condition is implemented using
transverse locality and it allows for bypassing the step of finding
initial conditions (counterterms) needed for differential equations, by
introducing a running coupling constant.  A question arises because
studies of asymptotically free matrix models \cite{GlazekWilson3}
\cite{PerrySzpigel} show that effective Hamiltonian matrices that are
suitable for bound state calculations may significantly deviate from a
self-replicating (fixed point) flow with one coupling constant.  But in
the case of many couplings helpful conditions may be provided by
coupling coherence \cite{BrisudovaPerryWilson}, which may work in the
approach of Eq.  (1.2) with a $P^+$-independent width parameter,
assuming that the cluster property breaking does not complicate the
coherence.  Kylin, Allen and Perry \cite{KylinAllenPerry} extended the
approach of Allen and Perry to massive particles.

The similarity renormalization group procedure for Hamiltonians is 
more flexible in defining its generator than Eqs. (1.1) and (1.2) 
might suggest. This flexibility is used to preserve boost symmetry
and cluster properties simultaneously.  Note that the boost invariance 
we talk about means invariance under boosts along the front and boosts 
corresponding to rotations about transverse axes in the infinite momentum 
frame. These boosts are sufficient for reaching states in all possible
uniform motions.

The desired boost-invariant operator formulation of similarity
\cite{GlazekAPP} is based on a transformation ${\cal U}_\lambda$ that
changes creation and annihilation operators of bare particles (denoted
by $q_\infty$, since they correspond to $\lambda = \infty$) into
creation and annihilation operators of effective particles corresponding
to the width $\lambda$ (denoted by $q_\lambda$).  Namely,

$$ q_\lambda \quad = \quad {\cal U}_\lambda \,\, q_\infty \,\, {\cal
U}^\dagger_\lambda \quad . \eqno(1.3a) $$

\noindent ${\cal U}_\lambda$ is secured to be unitary by construction.
Hamiltonian operators of all widths are assumed equal and when they are
expressed in terms of different creation and annihilation operators, the
coefficient functions change.  We have $H_\lambda(q_\lambda) =
H_\infty(q_\infty)$.  Assuming that Hamiltonians calculable in
perturbation theory contain only finite products of creation and
annihilation operators and applying the transformation ${\cal
U}_\lambda$, one obtains ${\cal H}_\lambda \equiv H_\lambda(q_\infty) =
{\cal U}^\dagger_\lambda H_\infty(q_\infty) {\cal U}_\lambda$.  This
relation means that the operator ${\cal H}_\lambda$ has the same
coefficient functions in front of products of $q_\infty$ as the
effective Hamiltonian $H_\lambda$ has in front of the unitarily
equivalent products of $q_\lambda$.  Differentiating ${\cal H}_\lambda$
one obtains

$$ {d\over d\lambda} \,\, {\cal H}_\lambda \quad = \quad - \quad [{\cal
T}_\lambda, \, {\cal H}_\lambda] \quad , \eqno (1.3b)$$

\noindent where the generator ${\cal T}_\lambda$ is related to ${\cal
U}_\lambda$ by

$$ {\cal T}_\lambda \quad = \quad {\cal U}^\dagger_\lambda \, {d\over
d\lambda} \, {\cal U}_\lambda \quad . \eqno(1.3c) $$

\noindent The script letters are introduced in Eqs.  (1.3a-c) to indicate 
that the operators can be conveniently thought about as expanded 
into sums of products of operators $q_\infty$.  The latter are 
independent of $\lambda$ and are not
differentiated in Eqs.  (1.3b-c).  In other words, Eqs.  (1.3b-c)
describe only the flow of coefficients in front of the creation and
annihilation operators.  Effective Hamiltonians are obtained from ${\cal
H}_\lambda$ using $ H_\lambda(q_\lambda) = {\cal U}_\lambda {\cal
H}_\lambda {\cal U}^\dagger_\lambda$.

Note that the operator ${\cal U}_\lambda$ is analogous to the Melosh
operator \cite{Melosh}.  The analogy stems from that the effective
particles corresponding to a small Hamiltonian width $\lambda$ can be
associated with constituent quarks or gluons from hadronic rest frame
spectroscopy while the bare particles with $\lambda \rightarrow \infty$
can be associated with current quarks and gluons from the infinite
momentum frame.  The boost-invariant similarity renormalization group
procedure for light-front Hamiltonians \cite{GlazekAPP} makes the
notions of current and constituent quarks and gluons boost invariant.
The procedure provides a calculable dynamical content to the
transformation that connects the current and constituent particles
independently of the reference frame one works in.  This may sound
confusing if Melosh transformation is associated with changing a frame
of reference. But the light-front dynamics is invariant with
respect to boosts in perturbative calculations and this property allows
for translating a Melosh-like transformation to one frame of reference,
where it simply relates bare particles and complex effective particles
in one theory in the same frame.

Boost invariance is guaranteed through the definition of the generator
${\cal T}_\lambda$.  Details will be reviewed in the next Section.  The
effective Hamiltonian $H_\lambda$ is given by a diagonal proximum of
certain operator $G_\lambda$, {\it i.e.} $H_\lambda =
F_\lambda[G_\lambda]$, while the generator ${\cal T}_\lambda$ is related
to a diagonal remotum of the same operator $G_\lambda$.  Following
\cite{GlazekAPP}, we introduce ${\cal G}_\lambda = {\cal
U}^\dagger_\lambda G_\lambda {\cal U}_\lambda$ and use

$$ [{\cal T}_\lambda, \, {\cal H}_{0\lambda}] \quad = \quad {d \over d
\lambda} \, (1 - F_\lambda)[{\cal G}_\lambda] \quad . \eqno(1.4) $$

\noindent The commutator structure guarantees that effective
Hamiltonians contain only connected interactions.

Now, the diagonal proximum can be defined in a boost-invariant way.  The
point is that the operation $F_\lambda$ multiplies every interaction
term in an effective Hamiltonian by a form factor which is a function of
boost invariant combinations of momenta labeling creation and
annihilation operators in that interaction term.  Such combinations can
include differences or sums of invariant masses.  The resulting
generator of the similarity transformation does not depend on the total
momentum of any state, contrary to Eq.  (1.2).  According to
\cite{GlazekAPP}, a necessary rescaling of the flow parameter, analogous
to rescaling provided by the total $P^+$ in Eq.  (1.2), is provided for
each Hamiltonian term separately using a parent three-momentum for that
term.  The parent momentum is defined as half of the sum of all momenta
labeling creation and annihilation operators in the
term in question.  All light-front Hamiltonians we consider are sums of
terms which contain integrals over the parent momenta.  Usually, when a
term acts on some state, a parent momentum equals only some fraction of
the total momentum of that state.  This fraction is defined by the
single interaction act and lies in a range of values allowed by momentum
conservation in the effective Schr\"odinger dynamics.  Thus, effective
interactions are independent of the total momenta of states they act on
and no dependence on spectators is generated.  Therefore, no cluster
property is violated in defining the similarity generator through Eq.
(1.4).  A considerable freedom is still left in choosing details of
$F_\lambda$.

In summary, the flexibility available in defining the similarity
transformation generator can be used to obtain a boost-invariant
band-diagonal structure of effective Hamiltonians preserving cluster
decomposition properties, through a suitable choice of the similarity
form factors $F_\lambda$.  This aspect of the procedure is the subject
of the present article, quite independently of which theory or
singularity is addressed.  Naturally, the effective theory is easiest to
make boost invariant if the initial Hamiltonian is regulated in a boost
invariant way.  For in that case the counterterms do not have to correct
the boost invariance breaking which a frame dependent regularization
would introduce.  A suitable class of regularizations will be described
in the next Section.

This paper is organized as follows.  Section 2 reviews necessary
elements of Ref.  \cite{GlazekAPP} set up for a third order calculation
of the running coupling constant in Hamiltonians of scalar $\phi^3$
theory in 6 dimensions.  The calculation is based on a plain power
series expansion in the coupling constant.  Counterterms are derived in
Section 3 and the effective coupling constant flow is calculated in
Section 4. Section 5 concludes the paper.  Appendix presents formulae
for third order effective vertex function with arbitrary momenta and
masses.

\vskip.3in
{\bf 2. EFFECTIVE HAMILTONIANS}
\vskip.1in
\nopagebreak

We focus our attention on a perturbative derivation of effective
Hamiltonians in scalar quantum field theory with an interaction term
$\sim \phi^3$ in 6 dimensions.  The theory is known to be asymptotically
free. The coupling constant in the initial regularized Hamiltonian is 
assumed to be infinitesimally small.  Our procedure for calculating effective
Hamiltonians closely follows Ref.  \cite{GlazekAPP}.

\vskip.3in
{\bf 2.1 Initial Hamiltonian}
\vskip.1in
\nopagebreak

The classical Lagrangian of the scalar field theory in question is

$$ {\cal L} = {1 \over 2} ( \partial_\mu \phi \partial^\mu \phi - \mu^2
\phi^2) - {g \over 3 !  } \phi^3 \quad . \eqno(2.1) $$

\noindent The corresponding light-front Hamiltonian reads \cite{Yan}
\cite{BrodskyRoskiesSuaya} ($x^\pm = x^0 \pm x^3$)

$$ H = \int dx^- d^4x^\perp \,\, \left[ {1 \over 2} \phi
(-\partial^{\perp \, 2} + \mu^2) \phi + {g \over 3 !  } \phi^3
\right]_{x^+ = 0} \quad . \eqno(2.2) $$

\noindent A quantum field $\phi(x)$ for $x^+=0$ is expanded in
its Fourier components,

$$ \phi(x)|_{x^+ = 0} = \int_{\delta^+} [k] \, ( a_k e^{-ikx} +
a^\dagger_k e^{ikx})|_{x^+ = 0} \quad , \eqno(2.3) $$

\noindent where the abbreviated notation for the integral means

$$ \int_{\delta^+} [k] \quad = \quad \int^\infty_{\delta^+} \, { dk^+
\over 2 k^+ } \int {d^4 k^\perp \over (2 \pi)^5 } \quad . \eqno(2.4) $$

\noindent The small parameter $\delta^+$ limits the longitudinal
momenta, $k^+$, from below.  The creation and annihilation operators
satisfy commutation relations

$$ [a_k, a^\dagger_q] = 2k^+(2\pi)^5 \delta(k^+ - q^+)\delta^4(k^\perp -
q^\perp) \quad , \eqno(2.5) $$

\noindent together with $ [a_k, a_q] = [a^\dagger_k, a^\dagger_q] = 0$.
Substituting Eq.  (2.3) in (2.2) one obtains the following unique
structure of the light-front Hamiltonian

$$ H = \int_{\delta^+}[k] {k^{\perp \, 2} + \mu^2 \over k^+} a^\dagger_k
a_k \,\, + \,\, {g \over 2} \int_{\delta^+} [kpq] \, 2(2\pi)^5
\delta^5(k+p-q)\,(a^\dagger_k a^\dagger_p a_q + a^\dagger_q a_p a_k) \quad
. \eqno(2.6) $$

\noindent It is unique in comparison to Hamiltonians in other forms of
dynamics because there are no terms that contain only creation or only
annihilation operators in Eq.  (2.6).  This feature is related to the
problem of ground state formation, since $H |0\rangle = 0$, where $a_k
|0\rangle = 0$, and only $a^\dagger_0 a^\dagger_0 a^\dagger_0$ and
$a_0^\dagger$ terms in $H$ could alter this feature, but they are absent
due to the cutoff $\delta^+$.  Readers interested in non-perturbative
aspects of Hamiltonian dynamics in the front form should consult
\cite{Susskind}.  However, no problems with modes of $k^+ = 0$ arise in
perturbation theory for Hamiltonians in the present paper.

The initial Hamiltonian for the similarity renormalization group
procedure is obtained from Eq.  (2.6) by introducing an additional
ultraviolet regularization factor (to be described below) and taking the
limit $\delta^+ \rightarrow 0$.  The initial Hamiltonian is denoted by
$H_\infty$, since it corresponds to the initial width $\lambda = \infty$
in the similarity flow, as discussed in the Introduction.  Creation and
annihilation operators that appear in $H_\infty$ were denoted in the
Introduction by $q_\infty$.

$H_\infty$ contains a regularization factor denoted by $r_\Delta$.
$\Delta$ stands for an ultraviolet cutoff parameter.  $r_\Delta \rightarrow
1$ when $\Delta \rightarrow \infty$.  In asymptotically free theories,
$\Delta$ can be sent to infinity when the renormalization process passes
the stage of calculating counterterms and deriving effective
Hamiltonians of finite width $\lambda$.  Still, the initial Hamiltonian
contains counterterms which depend on $r_\Delta$ and are denoted by
$X_\Delta$.  Besides diverging $\Delta$-dependent terms, $X_\Delta$
contains also finite parts that remove $\Delta$-independent
regularization effects caused by the factor $r_\Delta$.  Thus, we have

$$ H_\infty \quad = \quad \int [k] {k^{\perp \, 2} + \mu^2 \over k^+}
a^\dagger_{\infty k} a_{\infty k} \quad + $$
$$+ \quad {g \over 2} \int
[k_1 k_2 k_3] \, 2(2\pi)^5 \delta^5(k_1 + k_2 - k_3)\,(a^\dagger_{\infty k_1}
a^\dagger_{\infty k_2} a_{\infty k_3} + a^\dagger_{\infty k_3} a_{\infty
k_2} a_{\infty k_1} ) \, r_\Delta \,\, + \,\, X_\Delta \quad . $$
$$ \eqno(2.7) $$

\noindent The parameter $\delta^+$ is set equal 0 and this is why it is
not indicated, in distinction from Eq.  (2.6) where it was kept larger
than 0.

The regularization factor $r_\Delta$ has a simple form which results
from the following steps (applicable to the interaction term written out
explicitly in Eq.  (2.7) as well as to all counterterms $X_\Delta$
derivable in perturbation theory).  For a term containing a product of
$u$ creation and $v$ annihilation operators we define a parent momentum,
denoted by $P^+_{uv}$, which equals half of the sum of momenta labeling
all the operators.  For each momentum label $k_i$, with $i$ running
through $u + v$ values, we introduce $x_i = k_i^+/P^+_{uv}$ and
$\kappa^\perp_i = k^\perp_i - x_i P^\perp_{uv}$.  The regularization
factor is defined as

$$ r_\Delta \quad = \quad \prod_{i=1}^{u+v} \exp{- \eta_i \,
\kappa^{\perp \, 2}_i \over \Delta^2 } \quad , \eqno(2.8) $$

\noindent where $\eta_i = \eta(x_i)$ and $\eta$ is a useful function of
its argument.  One natural choice is $\eta(x) = 1$, for it is simple.
Another choice is $\eta(x) = 1/x$, a natural one because it appears in
invariant masses.  We shall assume the function $\eta$ lies between the
two choices but otherwise it will be left unspecified.  Leaving $\eta$
unspecified will help us identify finite regularization effects.

Consequently, the regularization factor in the second term in Eq.  (2.7)
is

$$ r_\Delta \quad = \quad \exp{- (\eta_1 + \eta_2) \, \kappa^{\perp \,
2}_{12} \over \Delta^2 } \quad , \eqno(2.9) $$

\noindent where $x_1=k_1^+/k_3^+$ and $\kappa_{12}^\perp = k_1^\perp -
x_1 k_3^\perp$.

The initial Hamiltonian of the similarity renormalization group
procedure in $\phi^3$ theory in 6 dimensions is given by Eq.  (2.7).  In
order to derive the third order running coupling constant in effective
Hamiltonians, we have to calculate $X_\Delta$ up to the terms order
$g^3$.  The counterterms are calculated order by order along the
evaluation of renormalization group flow for effective Hamiltonians.

\vskip.3in
{\bf 2.2 Similarity Flow of Hamiltonians}
\nopagebreak
\vskip.2in
\nopagebreak

The effective Hamiltonians are written as

$$ H_\lambda \quad = \quad F_\lambda [G_\lambda] \quad , \eqno(2.10) $$

\noindent where

$$ G_\lambda \quad = \quad {\cal U}_\lambda {\cal G}_\lambda {\cal
U}^\dagger_\lambda \quad . \eqno(2.11) $$

\noindent The operator ${\cal G}_\lambda$ is divided into two parts,
${\cal G}_\lambda = {\cal G}_0 + {\cal G}_{I\lambda}$, where ${\cal G}_0
= {\cal G}_\infty(g=0)$.  ${\cal G}_{I\lambda}$ satisfies the following
differential equation as a consequence of Eqs.  (1.3a)-(1.4) (see
Ref.  \cite{GlazekAPP})

$$ {d \over d\lambda} {\cal G}_{I\lambda} \quad = \quad \left[ f{\cal
G}_I, \, \left\{ {d \over d\lambda} (1-f){\cal G}_I \right\} _{{\cal
G}_0} \right] \quad . \eqno(2.12) $$

\noindent We dropped the subscript $\lambda$ on the right-hand side for
clarity.  $f$ denotes the similarity form factor introduced by
$F_\lambda$ and the curly bracket with the subscript ${\cal G}_0$
denotes a solution for $ {\cal T}_\lambda$ resulting from Eq.  (1.4).
We will omit the subscript ${\cal G}_0$ from now on and the Reader
should remember that a curly bracket implies an energy denominator for
every term in the bracket, {\it i.e.} a factor equal to inverse of the
eigenvalue of ${\cal G}_0$ corresponding to momentum labels of all
annihilation operators in a term minus eigenvalue of ${\cal G}_0$
corresponding to all creation operators in the term.

The similarity factor is defined for any operator of the form described
above Eq.  (2.8) as

$$ f_\lambda(u, v) \quad = \quad \exp{ [ - ({\cal M}^2_{u} - {\cal
M}^2_{v})^2/\lambda^4]} \quad . \eqno(2.13) $$

\noindent The script notation for invariant masses means 
$ {\cal M}^2_u = (k_1 + ...  + k_u)^2$, where the minus components
of the momentum four-vectors are given by $k^-_i = (k^{\perp\,2}_i +
\mu^2)/k^+_i$ for $i = 1, ..., u$.

Let us denote differentiation with respect to $\lambda$ by a prime
and expand the effective interaction in powers of the coupling constant
$g$ as

$$ {\cal G}_I \quad = \quad \sum_{n=1}^\infty \, \tau_n \quad ,
\eqno(2.14) $$

\noindent where $\tau_n$ denotes the sum of all terms order $g^n$ in
${\cal G}_\lambda$. Equation (2.12) implies

$$ \tau_n' \quad = \quad \sum_{k=1}^{n-1} \, [\tau_k, \,
\{(1-f)\tau_{n-k}\}'] \quad , \eqno(2.15) $$

\noindent and the first three terms in the expansion satisfy the
following equations

$$ \tau_1' \quad = \quad 0 \quad , \eqno(2.16a) $$

$$ \tau_2' \quad = \quad [\{f'\tau_1\}, f\tau_1] \quad , \eqno(2.16b) $$

$$ \tau_3' \quad = \quad [f\tau_1, \, \{-f'\tau_2 + (1-f)\tau_2'\}] \,\,
+ \,\, [f\tau_2, \,\{-f'\tau_1\}] \quad . \eqno(2.16c) $$

Equation (2.16a) implies that $\tau_1$ is independent of $\lambda$ and
equals the second term in the initial Hamiltonian from Eq.  (2.7).  In
other words, $\tau_{\lambda 1} = \tau_{\infty 1}$.  The corresponding
effective Hamiltonian interaction term is obtained by multiplying the
integrand in Eq.  (2.7) by $f_\lambda(12,3) = \exp{\{ - [(k_1+k_2)^2 -
\mu^2]^2/\lambda^4 \}}$ and transforming operators $a_{\infty k}$ into
$a_{\lambda k}$ using ${\cal U}_\lambda$ after the evaluation process
for ${\cal G}_\lambda$ is carried out to the desired order.

This last step is unusual in the sense that the operator ${\cal
U}_\lambda$ depends on the regularization (in scalar theory, the
dependence is reduced to ${\cal U}_\lambda$ being a functional of
$r_\Delta$) but, at the same time it is unitary and thus, it transforms
finite products of creation and annihilation operators by effectively
replacing everywhere $a_\infty$ by $a_\lambda$ and no other trace of
${\cal U}_\lambda$ is left in the effective Hamiltonian.  If ${\cal
G}_\lambda$ is found to order $g^n$, leaving terms $o(g^{n+1})$ still
undetermined, then ${\cal T}_\lambda$ is determined up to terms order
$g^n$ from Eq.  (1.4).  ${\cal T}_\lambda$ is antihermitian order by
order and determines a unitary ${\cal U}_\lambda$ to order $g^n$,
denoted ${\cal U}^{(n)}_\lambda$. Therefore,

$$ {\cal U}^{(n)}_\lambda \,\, \prod_{i \in u} a^\dagger_{\infty i} \prod_{j
\in v} a_{\infty j} \,\, {\cal U}^{(n)\dagger}_\lambda \quad = \quad \prod_{i
\in u} a^\dagger_{\lambda i} \prod_{j \in v} a_{\lambda j} + o(g^{n+1})
\quad , \eqno(2.17) $$

\noindent and the regularization dependence of ${\cal U}_\lambda$ does
not show up in $H_\lambda$ to order $g^n$ once ${\cal G}_\lambda$ is
renormalized to order $g^n$.  In successive orders, by construction, the
counterterms in ${\cal G}_\infty$ preserve unitarity of ${\cal
U}_\lambda$.  A perturbative proof of renormalizability for effective
Hamiltonians would require demonstration that there exists a set of
counterterms that remove regularization dependence from finite momentum
parts of ${\cal G}_\lambda$ to all orders of perturbation theory when
$\Delta \rightarrow \infty$ ({\it c.f.} \cite{GlazekWilson1}) and that
the resulting theory predicts covariant results.  The present article
does not demonstrate such set exists in the case of $\phi^3$ theory.
Also, our calculation is limited to terms order $g$, $g^2$ and $g^3$.
To verify that an effective Hamiltonian containing a running coupling
constant order $g^3$ may produce covariant results for scattering
processes, the present calculation must be extended to 4th order.

The transformation connecting two effective Hamiltonians with different
finite widths $\lambda_1$ and $\lambda_2$ is given by ${\cal
U}_{\lambda_1}{\cal U}^\dagger_{\lambda_2}$.  One can easily see that
the latter is free from dependence on $r_\Delta$ once ${\cal G}_\lambda$
is made independent of $r_\Delta$.  It is sufficient to observe that for
an infinitesimal difference between $\lambda_1$ and $\lambda_2$ the
effective transformation ${\cal U}_{\lambda_1}{\cal
U}^\dagger_{\lambda_2}$ is given by the similarity generator that is
expressed in terms of the $r_\Delta$-independent ${\cal G}_\lambda$.
Integrating the infinitesimal changes one obtains the same conclusion
for finite changes of the width.

Evaluation of $\tau_2$ and $\tau_3$ involves calculation of two
counterterms.  Both $\tau_2$ and $\tau_3$ are more complicated in
structure than $\tau_1$ and their evaluation requires new notation.
Namely,

$$ \tau_1 \quad = \quad \alpha_{21} \quad + \quad \alpha_{12} \quad ,
\eqno(2.18a) $$

$$ \tau_2 \quad = \quad \beta_{11} + \beta_{31} + \beta_{13} +
\beta_{22} \quad , \eqno(2.18b) $$

$$ \tau_3 \quad = \quad \gamma_{21} + \gamma_{12} + \gamma_{41} +
\gamma_{14} + \gamma_{32} + \gamma_{23} \quad . \eqno(2.18c) $$

\noindent Each term contains products of creation and annihilation
operators with fixed numbers of the operators in a product.  The first
subscript indicates the number of creation operators,
$a_\infty^\dagger$, and the second subscript denotes the number of
annihilation operators, $a_\infty$.  For all the terms, $\pi_{uv} =
\pi^\dagger_{vu}$.

Equations (2.16a-b) imply for second order terms the following relations
 \cite{GlazekAPP},

$$ \beta'_{31} \quad = \quad 2 f_2 [\alpha_{21} \alpha_{21}]_{31} \quad
, \eqno(2.19a) $$

$$ \beta'_{13} \quad = \quad 2 f_2 [\alpha_{12} \alpha_{12}]_{13} \quad
, \eqno(2.19b) $$

$$ \beta'_{11} \quad = \quad 2 f_2 [\alpha_{12} \alpha_{21}]_{11} \quad
, \eqno(2.19c) $$

$$ \beta'_{22} \quad = \quad f_2 [\alpha_{21} \alpha_{12} + 4
\alpha_{12} \alpha_{21} ]_{22} \quad . \eqno(2.19d) $$

\noindent The brackets mean replacement of products $a_i a^\dagger_j$ by
commutators $[a_i, a^\dagger_j]$.  The remaining products of operators
contain as many creation and annihilation operators as indicated by the
bracket subscripts, in the normal order, according to the same
convention as in Eqs.  (2.18a-c).  The factor $f_2$ depends on the
momenta labeling creation and annihilation operators.  In Eqs.
(2.19a-d), the brackets with operators involve integrals over
three-dimensional momentum labels of all creation and annihilation
operators and over loop momenta in loops that result from contractions
(in second order here only Eq.  (2.19c) for $\beta'_{11}$ contains a
loop integral).  The factor $f_2$ is understood to appear under the
integrals.  Symbolically, its structure appears in Eq.  (2.16b) and
reads $f_2 = \{f'\}f - f\{f'\}$.  The negative sign results from the
commutator in Eq.  (2.16b) that guarantees that only connected terms
appear in the effective interactions.  This is a general property of the
similarity procedure.

The factor $f_2$ is the only factor depending on $\lambda$ on the
right-hand side of Eqs.  (2.19a-d).  Therefore, solutions
are

$$ \beta_{\lambda 31} \quad = \quad 2 {\cal F}_{2\lambda} \,
[\alpha_{21} \alpha_{21}]_{31} \quad , \eqno(2.20a) $$

$$ \beta_{\lambda 13} \quad = \quad 2 {\cal F}_{2\lambda} \,
[\alpha_{12} \alpha_{12}]_{13} \quad , \eqno(2.20b) $$

$$ \beta_{\lambda 11} \quad = \quad 2 {\cal F}_{2\lambda} \,
[\alpha_{12} \alpha_{21}]_{11} \quad + \quad \beta_{\infty 11} \quad ,
\eqno(2.20c) $$

$$ \beta_{\lambda 22} \quad = \quad {\cal F}_{2\lambda} \, [\alpha_{21}
\alpha_{12} + 4 \alpha_{12} \alpha_{21} ]_{22} \quad , \eqno(2.20d) $$

\noindent where $\beta_{\infty 11}$ is a counterterm whose structure is
shown in the next Section to be

$$ \beta_{\infty 11} = \int [k] \, {\delta \mu^2_\infty \over k^+}
a^\dagger_{\infty k} a_{\infty k} \quad . \eqno(2.21) $$

The only new element of solutions (2.20a-d) which requires explanation
is the factor ${\cal F}_{2\lambda} = \int_\infty^\lambda
f_2$.  It is given by the following expression, which is a consequence
of Eq.  (2.13),

$$ {\cal F}_{2\lambda}(a, b, c) \quad = \quad { P^+_{ba} ba + P^+_{bc}
bc \over ba^2 + bc^2 } \, \left[f_\lambda(a,b) f_\lambda(b,c)-1\right]
\quad . \eqno(2.22) $$

\noindent Arguments $a$, $b$ and $c$ denote three successive momentum
configurations appearing in brackets in Eqs.  (2.19a-d) in the order
from the left to right, {\it i.e.} $a$ denotes the momenta labeling
creation operators in the brackets, $c$ denotes the momenta labeling
annihilation operators and $b$ denotes the intermediate configuration,
which includes the momenta labeling operators contracted in the brackets
and momenta labeling creation operators originating from the interaction
connecting configuration $b$ with $c$ and momenta labeling annihilation
operators originating in the interaction connecting configuration $a$
with $b$.

$P^+_{uv}$ denotes parent momentum for the whole connected interaction
sequence between momentum configurations $u$ and $v$ (in the second
order case here the sequence reduces to a single interaction vertex
order $g$, but the definition of $P^+_{uv}$ remains valid in higher
order cases later).  The symbols $ba$ and $bc$ denote differences of
invariant masses, as explained below.  We use abbreviations $ba^2 \equiv
(ba)^2$ {\it etc.}

For any two momentum configurations $u$ and $v$

$$ uv = {\cal M}^2_{uv} - {\cal M}^2_{vu} \quad , \eqno(2.23a) $$

\noindent where

$$ {\cal M}^2_{uv} = \left[ \sum_{i \in u(v)} k_i \right]^2 \eqno(2.23b)
$$

\noindent and $u(v)$ denotes those momenta from the configuration $u$
that are involved in interactions acting between the configurations $u$
and $v$.  Minus components of all momenta are given by $k^- = (k^{\perp
\, 2} + \mu^2)/k^+$.  As an example of this notation, the similarity
form factor from Eq.  (2.13) reads

$$ f_\lambda(u,v) = \exp{ - vu^2 \over \lambda^4 } \quad . \eqno(2.23c)
$$

Equation (2.16c) for third order interactions implies the following
result for terms relevant to the running coupling constant evaluation,

\begin{eqnarray*}
&& \gamma'_{21} \quad = \quad f_3 \, \left[ \, 8 \left[\alpha_{12}
\alpha_{21} \alpha_{21}\right]_{21} \, + \, 4 \left[ [\alpha_{12}
\alpha_{21}] \alpha_{21} \right]_{21} \, + \,
2 \left[ \alpha_{21} [\alpha_{12} \alpha_{21}]\right]_{21}
\, \right] \, + \\
&& \\
&& - \, 2 \{f'\} \left[ \beta_{\infty 11} \alpha_{21} \right]_{21} \, +
\, \{f'\}\left[ \alpha_{21} \beta_{\infty 11} \right]_{21} \quad .
\end{eqnarray*}
$$\eqno(2.24) $$

\noindent The factor $f_3$ has the structure $f_3 \, = \, [f\{(1-f){\cal
F}_2\}' \, + \, \{f'\}f{\cal F}_2 ] \, - \, [\{(1-f) {\cal F}_2 \}'f \,
+ \, f {\cal F}_2 \{f'\}]$.  Integration of both sides of Eq.  (2.24)
gives

\begin{eqnarray*} && \gamma_{\lambda 21} \quad = \quad {\cal
F}_{3\lambda} \, \left[ \, 8 \left[\alpha_{12} \alpha_{21}
\alpha_{21}\right]_{21} \, + \, 4 \left[ [\alpha_{12} \alpha_{21}]
\alpha_{21} \right]_{21} \, + \, 2 \left[ \alpha_{21} [\alpha_{12}
\alpha_{21}] \right]_{21} \right] \, + \\ && \\ && + \, 2
\{1-f_\lambda\} \left[ \beta_{\infty 11} \alpha_{21} \right]_{21} \, -
\, \{1-f_\lambda\}\left[ \alpha_{21} \beta_{\infty 11} \right]_{21} \, +
\, \gamma_{\infty 21} \quad , \end{eqnarray*}
$$ \eqno(2.25) $$

\noindent where $\gamma_{\infty 2 1}$ denotes the third order
counterterm, to be calculated in the next Section.  A new element of Eq.
(2.25) is the factor ${\cal F}_{3\lambda} = \int_\infty^\lambda f_3$.
It appears in front of operator brackets that involve four successive
momentum configurations denoted from the left to right by $a$, $b$, $c$
and $d$.  The brackets contain one loop integral.  The configuration $a$
has two momenta, $k_1$ and $k_2$, while the configuration $d$ only one,
$k_3$ [{\it c.f.} Eq.  (2.7)].  We have

$$ {\cal F}_{3\lambda}(a,b,c,d) \quad = \quad F_3(a,b,c,d) +
F_3(d,c,b,a) \quad , \eqno(2.26a) $$

\noindent where

\begin{eqnarray*}
&&F_3(a,b,c,d) = {P^+_{cb}cb + P^+_{cd}cd \over cb^2 + cd^2}
\left[(P^+_{bd}bd + P^+_{ba}ba)\left[
{f_{ab}f_{bc}f_{cd}f_{bd} - 1 \over ab^2 + bc^2 + cd^2 + bd^2} -
{f_{ab}f_{bd} - 1 \over ab^2 + bd^2}\right] + \right. \\
&& \\
&&\ + \left. P^+_{bd} {bc^2 + cd^2 \over db}
\left[ {f_{ab}f_{bc}f_{cd} - 1 \over ab^2 + bc^2 + cd^2} -
{f_{ab}f_{bc}f_{cd}f_{bd} - 1 \over ab^2 + bc^2 + cd^2 + bd^2}\right]
\right] \, ,
\end{eqnarray*}
$$ \eqno(2.26b)$$

\noindent and $f_{ab}$ is an abbreviated notation for $f_\lambda(a,b) =
\exp{[-ab^2/\lambda^4]}$; {\it c.f.} Eq.  (2.23c) above.

\vskip.3in
{\bf 3. COUNTERTERMS THROUGH ORDER $g^3$}
\nopagebreak
\vskip.1in
\nopagebreak

The two counterterms, $\beta_{\infty 1 1}$ from Eq.  (2.20c) and
$\gamma_{\infty 2 1}$ from Eq.  (2.25), are calculated using the
equations they appear in.  The counterterms are determined by the
condition that those equations are independent of $r_\Delta$ when
$\Delta \rightarrow \infty$ for arbitrary finite values of $\lambda$ and
particle momenta.  We first describe calculation of $\beta_{\infty 1 1}$
and then $\gamma_{\infty 2 1}$.

\vskip.3in
{\bf 3.1 Mass counterterm}
\nopagebreak
\vskip.1in
\nopagebreak

Equation (2.20c) implies

$$ \beta_{\lambda 11} = \int [k] \, {\delta \mu^2_\lambda \over k^+}
a^\dagger_{\infty k} a_{\infty k} \quad , \eqno(3.1) $$

\noindent where

$$ \delta \mu^2_\lambda = \delta \mu^2_\infty + \left({g \over
2}\right)^2 {1\over 2 (2\pi)^5} \int_0^1 {dx \over x(1-x) } \int d^4
\kappa^\perp {2 \over {\cal M}^2 - \mu^2} \left[ f^2_\lambda ({\cal
M}^2, \mu^2) - 1\right] \, r_{\Delta \beta}\quad . \eqno(3.2a) $$

\noindent The script ${\cal M}$ denotes invariant mass, ${\cal M}^2 =
(\kappa^{\perp \, 2} + \mu^2)/x(1-x)$, and the regularization factor
comes out to be

$$ r_{\Delta \beta} \quad = \quad \exp{ \left\{-2[\eta(x) + 
\eta(1-x)]\kappa^{\perp \,2}/\Delta^2 \right\} } \quad . \eqno(3.2b) $$

\noindent $\delta \mu^2_\infty$ in Eq.  (3.2a) is the counterterm
contribution.  The counterterm is of the form given in Eq.  (2.21) since
the integral in the second term in Eq.  (3.2a) is independent of the
momentum $k$ and depends on regularization.  The result of integration
depends on the cutoff parameter $\Delta$ and function $\eta$.  A finite,
regularization dependent part of the result remains undetermined.
Therefore, we have to adjust its value to data.

Without loss of generality, we assume that some {\it gedanken} data
requires the mass squared parameter in effective Hamiltonian with
$\lambda = \lambda_0$ to be equal $\mu^2 + \delta \mu^2_0$.  Hence,

$$ \delta \mu^2_\infty = \delta \mu^2_0 - \left({g \over 2}\right)^2
{1\over 2 (2\pi)^5} \int_0^1 {dx \over x(1-x) } \int d^4 \kappa^\perp {2
\over {\cal M}^2 - \mu^2} \left[ f^2_{\lambda_0} ({\cal M}^2, \mu^2) -
1\right] \, r_{\Delta \beta}\quad . \eqno(3.3) $$

\noindent Integration gives two diverging terms, one proportional to
$\Delta^2$ and another one proportional to $\log{\Delta}$.  The
remaining finite part depends on our choice of the function $\eta$.
Evaluating the integral for $\eta(x) = 1/x$ one obtains

$$ \delta \mu^2_\infty = g^2 {1 \over (4\pi)^3} \left[ \, {1\over
24}\Delta^2 \, - \, \mu^2 {5\over 6} \log{\Delta \over \mu} \, + \,
\mu_\eta^2 \right] \quad , \eqno(3.4) $$

\noindent where $\mu_\eta$ has a finite limit when $\Delta \rightarrow
\infty$.  The logarithmically divergent part is independent of the
function $\eta$ and agrees with results for the Lagrangian mass squared
counterterm obtained using Feynman diagrams and dimensional
regularization \cite{MacfarlaneWoo} \cite{Collins} in the following
sense:  when one changes $\Delta$ to $\Delta'$ the logarithmic part of
the counterterm changes with $\Delta$ as the mass squared changes as a
function of the renormalization scale in Eq.  (7.1.22) in
\cite{Collins}.

The resulting mass squared term in the effective Hamiltonian can be
written in the limit $\Delta \rightarrow \infty$ as

\begin{eqnarray*}
&&  \mu_\lambda^2 \quad = \quad \mu^2 \, + \, \delta \mu^2_\lambda \quad
= \\
&& \\
&& = \mu^2 + \delta \mu^2_0 + \left({g \over 2}\right)^2 {1\over 2
(2\pi)^5} \int_0^1 {dx \over x(1-x) } \int d^4 \kappa^\perp {2 \over
{\cal M}^2 - \mu^2} \left[ f^2_\lambda ({\cal M}^2, \mu^2) -
f^2_{\lambda_0} ({\cal M}^2, \mu^2) \right] \quad .
\end{eqnarray*}
$$\eqno(3.5a) $$

\noindent The above result is particularly simple for $\mu = 0$ and in
that case it reads ($\delta \mu^2_0$ is proportional to $g^2$)

$$ \mu_\lambda^2 \quad = \quad \delta \mu^2_0 \, + \, g^2 {1\over
(4\pi)^3} \, {1\over 24} \, \sqrt{\pi \over 2} \, (\lambda^2 -
\lambda_0^2) \quad . \eqno(3.5b) $$

\noindent Logarithmic dependence on $\lambda$ arises for $\mu > 0$.  The
value of $\delta \mu^2_0$ could be found, for example, by solving a
single physical meson eigenvalue problem, expressing the physical meson
mass in terms of $\delta \mu^2_0$ and adjusting the latter to obtain the
{\it gedanken} experimental mass value for mesons.

\vskip.3in
{\bf 3.2 Coupling constant counterterm}
\nopagebreak
\vskip.1in
\nopagebreak

The coupling constant counterterm $\gamma_{\infty 2 1}$ is evaluated
from Eq.  (2.25).  The mass squared counterterms cancel quadratic and
part of logarithmic divergences so that the vertex counterterm is needed
to remove only logarithmically divergent and finite regularization
effects.

The interaction term $\gamma_{\lambda 2 1}$ has the form

$$ \gamma_{\lambda 2 1} = \int [k_1 k_2 k_3] \, 2(2\pi)^5 \delta^5 (k_1
+ k_2 - k_3) \, \left[ \gamma_\lambda(k_1, k_2, k_3) + \gamma_\infty
(k_1, k_2, k_3) \right] \, a^\dagger_{\infty k_1} a^\dagger_{\infty k_2}
a_{\infty k_3} \, r_\Delta \quad . \eqno(3.6) $$

\noindent The counterterm function $\gamma_\infty(k_1, k_2, k_3)$
originates from $\gamma_{\infty 2 1}$ in Eq.  (2.25).  Since the
similarity renormalization group procedure preserves canonical
symmetries of light-front Hamiltonians, the functions
$\gamma_\lambda(k_1, k_2, k_3)$ and $\gamma_\infty (k_1, k_2, k_3)$
depend only on variables $x_1$ and $\kappa^\perp_{12}$, introduced in
Eq.  (2.9).

The entire regularization dependence of $ \gamma_\lambda(k_1, k_2, k_3)
\equiv \gamma_\lambda(x_1, \kappa^\perp_{12})$ in the limit $\Delta
\rightarrow \infty$ is contained in

\begin{eqnarray*}
&& \gamma_\lambda(x_1, \kappa^\perp_{12})|_{r_{\Delta}} \quad \equiv
\quad \gamma_\lambda(x_1, 0^\perp)|_{r_{\Delta}} \quad = \quad
\left({g \over 2}\right)^3 {\pi^2 \over 2(2\pi)^5} \times \\
&& \\
&& \times \left[ {1\over 2} \left[ \int_{x_1}^1 {dx \over x(1-x)(x-x_1)}
\int_0^\infty \kappa^2 d\kappa^2 \,\, 8 {x-x_1 \over x x_2 {\cal M}^4 }
\exp{ \left({- c_\eta \kappa^2 \over \Delta^2}\right)} + ( x_1
\leftrightarrow x_2 ) \right] + \right. \\
&& \\
&& \left. + \int_0^1 {dx \over x(1-x)} \int_0^\infty \kappa^2 d\kappa^2
{-3 \over {\cal M}^4} \exp{ \left({- d_\eta \kappa^2
 \over \Delta^2}\right)}\right] \quad ,
\end{eqnarray*} 
\nopagebreak
$$\eqno(3.7a)$$

\noindent where

$$ c_\eta = \eta (x) + \eta (1-x) + \left\{ \eta (x_1/x) + \eta
[(x-x_1)/x] \right\} (x_1/x)^2 + \eta[(x-x_1)/x_2] + \eta[(1-x)/x_2] \,
 \eqno(3.7b) $$

\noindent and

$$ d_\eta \quad = \quad 2 \, [\eta(x) + \eta(1-x)] \quad . \eqno(3.7c)$$

\noindent The first term in Eq.  (3.7a), symmetrized in $x_1$ and $x_2$,
originates from the first 8 terms in the long bracket in Eq.  (2.25).
The second term originates from the next 4 + 2 terms in the long bracket
and 2 + 1 mass counterterm terms in Eq.  (2.25).  The regularization
effects are independent of $\kappa^\perp_{12}$.  The apparent
singularity at $\kappa^2 \rightarrow 0$ for $\mu = 0$ is irrelevant to
the regularization dependence and appears here only because we do not
display similarity factors that remove the singularity.  The full
expression for $\gamma_\lambda(x_1, \kappa^\perp_{12})$ is given in
Appendix.

Dropping all parts that are independent of regularization, Eq.  (3.7a)
gives

\begin{eqnarray*}
&&  \gamma_\lambda(x_1, 0^\perp)|_{r_\Delta} \quad =
\quad \left({g \over 2}\right)^3 {1 \over (4\pi)^3} \times \\
&& \\
&& \left[ 3 \log{\Delta \over \mu} \,
- \, 4 \left[ \int_{x_1}^1 dx {1-x \over x_2} \log{c_\eta}
\, + \, (x_1 \leftrightarrow x_2) \right]
\, + \, 3 \int_0^1 dx \, x(1-x)
\log{d_\eta} \, \right] \quad .
\end{eqnarray*}
$$\eqno(3.8) $$

\noindent The bare boson mass, $\mu$, can be replaced by a finite
constant of the same dimension in the case of massless bosons.

Equation (3.8) says that the diverging regularization dependence of the
interaction vertex, {\it i.e.} the term proportional to $\log {\Delta}$,
is independent of the particle momenta and one can remove the divergence
by merely changing the initial coupling constant $g$ in Eq.  (2.7).
Thus, no diverging $x$-dependent counterterms are required - different 
situation than in \cite{overlap}.  
However, it is visible that the vertex contains a
finite regularization dependent part that is a function of $x_1$.  The
function depends on our choice for $\eta$.  One could completely
subtract the $\eta$-dependence for arbitrary choices of $\eta$ by
defining a counterterm that contains a negative of the $\eta$-dependent
part of the right-hand side of Eq.  (3.8).  For example, if $\eta = 1$
one has $c_\eta = 4 + 2(x_1/x)^2$ and $d_\eta = 4$.  Integration
produces a smooth and slowly varying function of $x_1$, which needs to
be subtracted.

Since the whole regularization effect is independent of $\lambda$ and
$\kappa^\perp_{12}$, it can be completely removed from
$\gamma_{\lambda}(x_1, \kappa_{12}^\perp)$ by subtracting
$\gamma_{\lambda_0}(x_1, 0^\perp)$, where $\lambda_0$ is chosen
arbitrarily.  However, one has to add back the finite regularization
independent part of $\gamma_{\lambda_0}(x_1, 0^\perp)$, which will be
denoted below by $\gamma_0(x_1)$.  The function $\gamma_0(x_1)$ is
necessary to recover Poincar\'e symmetry of observables.  Regularization
spoils Poincar\'e symmetry.  The symmetry may be restored once
counterterms remove regularization effects, but one is not allowed to
change terms independent of regularization.  Therefore, the function
$\gamma_0(x_1)$ must be reinserted.  This function is not altered when
$\lambda$ changes and could be considered marginal in analogy with usual
renormalization group analysis.

Although one can isolate finite $\eta$-dependent functions of $x_1$ in
the effective vertex, for the particular choice of the regularization
factors we adopt, the ultimate adjustment of the function
$\gamma_0(x_1)$ has to be delayed until 4th order calculations are
completed.  For there exists in $\phi^3$ theory no 3rd order scattering
amplitude one could use to find out what function $\gamma_0(x_1)$ 
renders Poincar\'e symmetry of scattering observables with our choice of
$r_\Delta$ in Eq.  (2.7).  It will be interesting to see if the explicit
dependence on $\eta$ isolates the same function as required by the
symmetry.  However, it should be pointed out that the function does not
influence the way the 3rd order running coupling constant in effective
Hamiltonians depends on $\lambda$.

The counterterm function $ \gamma_\infty (k_1, k_2, k_3) \equiv
\gamma_\infty (x_1)$, which removes the regularization dependence from
the effective vertex reads

$$ \gamma_\infty (x_1) \quad = \quad - \,\, \gamma_{\lambda_0}(x_1,
0^\perp ) \,\, + \,\, \gamma_0(x_1) \quad . \eqno(3.9a) $$

\noindent It can be used to define a new regularization dependent
coupling constant $g_\Delta$ in the initial Hamiltonian in Eq.  (2.7).
We select a convenient value of $x_1 = x_0$ and obtain

$$ {g_\Delta \over 2} \quad = \quad {g \over 2} \, + \,
\gamma_\infty(x_0) \quad = \quad {g \over 2} \, - \,
\gamma_{\lambda_0}(x_0, 0^\perp) + \gamma_0 \quad , \eqno(3.9b) $$

\noindent where $\gamma_0 \equiv \gamma_0(x_0)$.  Using Eq.  (3.8), we
see that the initial coupling $g$ is replaced by the new
$\Delta$-dependent quantity

$$ g_\Delta = g \left[ 1 - g^2 {3 \over 4(4\pi)^3 } \log{\Delta \over
m_0} \right] + o(g^5) \quad . \eqno(3.9c) $$

\noindent with certain constant $m_0$.  Thus, the theory exhibits
asymptotic freedom in 3rd order terms.  Our result agrees with
literature, say Eq.  (7.1.26) from \cite{Collins}, in the sense that
when we change $\Delta$, the change required in the coupling constant in
the initial Hamiltonian for obtaining $\Delta$-independent effective
Hamiltonians matches the change implied by Feynman diagrams and
dimensional regularization.  Comparison with Feynman calculus will be
farther discussed below.

Having derived the structure of counterterms we can proceed to
evaluation of the finite similarity flow of effective Hamiltonians
towards small widths $\lambda$.

\vskip.3in
{\bf 4. RUNNING COUPLING CONSTANT}
\vskip.1in
\nopagebreak

Our procedure for evaluating the running coupling constant in
Hamiltonians follows theory from \cite{GlazekWilson1} and
\cite{GlazekWilson2} using \cite{GlazekAPP}.  The procedure has been
outlined in a matrix model example in \cite{GlazekWilson3}.  Here, we
use particle creation and annihilation operators in a boost invariant
way, instead of matrix elements.

The initial Hamiltonian interaction vertex depends on regularization
through the factor $r_\Delta$ and corresponding counterterm.  The bare
coupling constant $g$ is replaced by the coupling constant $g_\Delta$
according to Eqs.  (3.9a-c), {\it i.e.}

$$ g_\Delta = g - 2 \left[ \gamma_{\lambda_0}(x_0,0^\perp) - \gamma_0
\right] \quad . \eqno(4.1) $$

\noindent The factor 2 is needed because $g/2$ appears in the
Hamiltonian.  Both $\gamma_{\lambda_0}(x_0)$ and $\gamma_0$ are
proportional to $g^3$.  Inverting the series (4.1) one can express $g$
in terms of $g_\Delta$.

Evaluation of $\gamma_{\lambda 2 1}$ in Eq.  (3.6) leads now to a finite
expression, which has a limit when $\Delta \rightarrow \infty$.  In that
limit, the effective Hamiltonian interaction term takes the form

$$ H_{\lambda 2 1} = \int [k_1 k_2 k_3] \, 2(2\pi)^5 \delta^5 (k_1 + k_2
- k_3) \,\, f_\lambda[(k_1 + k_2)^2, k_3^2] \,\,
V_\lambda(x_1,\kappa^\perp_{12} ) \,\, a^\dagger_{\lambda k_1}
a^\dagger_{\lambda k_2} a_{\lambda k_3} \quad , \eqno(4.2a) $$

\noindent where

$$ V_\lambda(x_1, \kappa^\perp_{12}) = {g_\Delta \over 2} +
\gamma_\lambda (x_1, \kappa^\perp_{12}) + \gamma_{\lambda_0} (x_0, 0^\perp)
- \gamma_{\lambda_0} (x_1, 0^\perp)]  + \gamma_0(x_1) - \gamma_0
\quad , \eqno(4.2b) $$

\noindent is the effective vertex function and $f_\lambda$ is the
similarity vertex form factor.  $\gamma_\lambda (x_1,
\kappa^\perp_{12})$ and the remaining terms in $V_\lambda(x_1,
\kappa^\perp_{12})$ are proportional to $g^3$.  Our calculation is done
to this order of accuracy only, so that $g^3 \equiv g_\Delta^3$ and Eq.
(4.2b) is considered to be an expansion in powers of $g_\Delta$.

We define the running coupling constant as the value of $2V_\lambda(x_1,
\kappa^\perp_{12})$ at a chosen configuration of momentum variables, 
denoted by $(x_{10}, \kappa^\perp_{120})$ and specified later. In other 
words, $g_\lambda = 2V_\lambda (x_{10}, \kappa^\perp_{120})$.  
This is a natural definition, analogous to the
standard Thomson limit in the case of electron charge in QED.  We have

$$ g_\lambda = g_\Delta + 2 [ \gamma_\lambda(x_{10},\kappa^\perp_{120})
+ \gamma_{\lambda_0}(x_0, 0^\perp) - \gamma_{\lambda_0}( x_{10},0^\perp)
+ \gamma_0(x_{10}) - \gamma_0 ] \quad . \eqno(4.3a) $$

\noindent It is natural to use $x_0 = x_{10}$.  Then,

$$ g_\lambda \quad = \quad g_\Delta \,\, + \,\, 2 \,
\gamma_\lambda(x_{10},\kappa^\perp_{120}) \quad . \eqno(4.3b) $$

\noindent This equation demonstrates that the effective coupling
constant $g_\lambda$ depends on the value of the finite function
$\gamma_0(x_1)$ in the counterterm at one point, the same as the one
used to define $g_\Delta$.

Suppose that for the chosen value of $\lambda = \lambda_0$, the running
coupling constant should have the value $g_{\lambda_0} = g_0$,
determined from comparison with experiment.  Then,

$$ g_0 \quad = \quad g_\Delta \,\, + \,\, 2
\gamma_{\lambda_0}(x_{10},\kappa^\perp_{120}) + o(g_\Delta^5) \quad ,
\eqno(4.4) $$

\noindent where $g$ in terms order $g^3$ is replaced by $g_\Delta$.
Inverting this series expansion we obtain

$$ g_\Delta \quad = \quad g_0 \,\, - \,\, 2 \gamma_{\lambda_0}
(x_{10},\kappa^\perp_{120}) + o(g_0^5) \quad , \eqno(4.5) $$

\noindent where in terms order $g_\Delta^3$ we have $g_\Delta$ replaced
by $g_0$.  Relation (4.5) can be inserted into Eq.  (4.3b) to yield

$$ g_\lambda \quad = \quad g_0 \,\, + \,\, 
2 \left[ \gamma_\lambda(x_{10},\kappa^\perp_{120})
- \gamma_{\lambda_0}( x_{10},\kappa^\perp_{120} ) \right] \,\,
+ \,\, o(g_0^5) \quad . \eqno(4.6) $$

\noindent This relation is free from dependence on the finite function
$\gamma_0(x_1)$ in the counterterm.

The vertex function in the effective interaction in Eqs.  (4.2a-b) is
equal

$$ V_\lambda(x_1, \kappa^\perp_{12}) = {g_0 \over 2} +
\gamma_\lambda(x_1, \kappa^\perp_{12}) -
\gamma_{\lambda_0}(x_{10},\kappa^\perp_{120}) +
\gamma_{\lambda_0}(x_{10}, 0^\perp) - \gamma_{\lambda_0}(x_1,0^\perp) +
\gamma_0(x_1) - \gamma_0 + o(g_0^5) \quad , \eqno(4.7a) $$

\noindent where in terms order $g^3$ the initial $g$ is replaced by
$g_0$.  The difference between $g$ and $g_0$ is of order $g_0^3$ and it
is included in terms $o(g_0^5)$.  Equation (4.7a) gives us the effective
vertex function for width $\lambda$ expanded in powers of the effective
coupling constant $g_0$ corresponding to width $\lambda_0$.  Written as
a power series in $g_\lambda$, the effective vertex function reads

$$ V_\lambda(x_1, \kappa^\perp_{12}) = {g_\lambda \over 2} +
\gamma_\lambda(x_1, \kappa^\perp_{12}) -
\gamma_{\lambda}(x_{10},\kappa^\perp_{120}) + \gamma_{\lambda_0}(x_{10},
0^\perp) - \gamma_{\lambda_0}(x_1,0^\perp) + \gamma_0(x_1) - \gamma_0 +
o(g_\lambda^5) \quad , \eqno(4.7b) $$

\noindent where in terms order $g^3$ one replaces $g$ by $g_\lambda$.
Clearly, Eq.  (4.7b) reproduces the relation $g_\lambda = 2
V_\lambda(x_{10}, \kappa^\perp_{120})$.

It remains to calculate the dependence of $g_\lambda$ on $\lambda$,
which requires a choice of the momentum configuration $(x_{10},
\kappa^\perp_{120})$.  The particular choice we will adopt is suitable
for massless bare bosons, {\it i.e.} for $\mu = 0$.  Our choice would
require a change for $\mu > 0$, to avoid vanishing of similarity factors
when $\mu^2/x_{10}$ tends to infinity.  But the change is not
significant since $\mu > 0$ introduces no alteration in our procedure
apart from the change of momentum configuration, and it also does not
interfere with boost-invariance.  Therefore, we will limit details of
our presentation to the simplest massless case, using a momentum
configuration that is most convenient when $\mu=0$.  We also continue to
keep $x_0$ in Eq.  (3.9b) equal to $x_{10}$.  For $\mu > 0$ one has to
make $x_0$ greater than 0, too.  Other choices of the momentum
configurations than we adopt here are equally possible in the case 
$\mu = 0$.  The one we
choose is particularly suitable for executing integrals over the loop
momenta and extracting the running coupling dependence on the width
$\lambda$ analytically.  Expressions for $\gamma_\lambda(x_1,
\kappa^\perp_{12})$ for $\mu \ge 0$ are given in Appendix.

In the massless case, the most convenient configuration is $x_{10} = 0$
and $\kappa^\perp_{120} = 0^\perp$.  For massive particles one considers
$1 > x_{10} > 0$ but there is no compelling reason to consider
$\kappa^\perp_{120} \neq 0$ in $\phi^3$ theory even for massive
particles.  Note that the parent momentum $k_3$ in Eq.  (4.2) is
arbitrary and not limited by our choice of the momentum configuration
$(x_{10}, \kappa^\perp_{120})$.  This feature is not readily available 
when one considers only specific matrix elements of effective Hamiltonians 
or when explicit boost invariance is not preserved by the renormalization 
group procedure.

Eq.  (4.6) gives

\begin{eqnarray*}
&& g_\lambda \quad = \quad g_0 + \\
&& \\
&& + g_0^3 {1\over 24} {1 \over (4\pi)^3}
\int_0^\infty
{dz\over z} \left[ 2(f_\lambda - f_\lambda^3) - 2(f_0 - f_0^3) +
20(f_\lambda^3 - f_\lambda^2) - 20(f_0^3 - f_0^2) + 9(f_0^2 -
f_\lambda^2)\right] \quad ,
\end{eqnarray*}
$$\eqno(4.8) $$

\noindent where $f_\lambda = \exp{ - z^2/\lambda^4}$ and $f_0 = \exp{ -
z^2/\lambda_0^4}$.  A straightforward integration gives

$$ g_\lambda \quad = \quad g_0 \,\, - \,\, g_0^3 \,\, { 3 \over 4(4\pi)^3}
\, \log{\lambda \over \lambda_0} \quad , \eqno(4.9) $$

\noindent which exhibits asymptotic freedom.  Differentiating with
respect to $\lambda$ and keeping terms up to order $g_\lambda^3$ one
obtains

$$ {d \over d \lambda} \,g_\lambda \quad = \quad - \quad g^3_\lambda
\,\, {3 \over 256 \pi^3} \,\, {1\over \lambda} \quad . \eqno(4.10) $$

\noindent This equation demonstrates the same $\beta$ function for
coupling constants in effective Hamiltonians as obtained in Lagrangian
approaches using Feynman diagrams and dimensional regularization, when
one identifies the renormalization scale with the Hamiltonian width
$\lambda$.  This is encouraging but one needs to remember that for
comparison of perturbative scattering amplitudes in Hamiltonian and
Lagrangian approaches it is necessary to make additional calculations
and at least of fourth order in $g$.  Beyond model matrix studies such
as in \cite{GlazekWilson3}, 4th order similarity calculations have so
far been carried out only in simplified Yukawa model by Mas{\l}owski and
Wi\c eckowski \cite{MaslowskiWieckowski}.

Integrating Eq.  (4.10) one obtains ($\alpha = g^2/4\pi$)

$$ \alpha_\lambda = { \alpha_0 \over 1 + \alpha_0 (3/32\pi^2)
\log{\lambda / \lambda_0} } \quad , \eqno(4.11) $$

\noindent which shows our result for a boost invariant running coupling
constant in effective Hamiltonians.  Our procedure explains how the
running coupling constant can be included in quantum mechanics of
effective particles, which is given by the Schr\"odinger equation with
the corresponding Hamiltonian $H_\lambda$.

\vskip.3in
{\bf 5. CONCLUSION}
\nopagebreak
\vskip.1in
\nopagebreak

A remarkable result of the whole procedure in the case of asymptotically
free scalar dynamics is an extremely simple structure of diverging
counterterms and a complete control over involved effective
interactions, in perturbation theory.  Our operator calculus preserves
cluster properties and allows for evaluation of effective Hamiltonians
without limitation to any particular set of matrix elements.  In other
words, we can easily derive integral expressions for matrix elements of
effective Hamiltonians in the whole Fock space spanned by basis states
of effective particles.  The renormalization group equations are
integrated analytically using gaussian similarity form factors.  Also,
the unitary nature of the similarity transformation for effective
particle creation and annihilation operators removes wave function
renormalization constant from the procedure. However, the regularization 
factor introduced in the initial Hamiltonian requires an additional
finite counterterm that contains a boost invariant function of 
longitudinal momentum fractions.

The renormalization group flow for Hamiltonians differs from standard
procedures applied to S-matrix.  The key difference from standard
procedures is that the running coupling constant is derived as a
coefficient in front of a certain term in an effective renormalized
light-front Hamiltonian instead of a Lagrangian term or in a scattering
amplitude.  The advantage of knowing an effective Hamiltonian is that
not only one can attempt to describe scattering processes but also bound
states using the corresponding eigenvalue equation.

Note that the light-front form of relativistic quantum field dynamics
re-defines the vacuum problem in a way that is only partly understood.
\cite{Wilsonetal} \cite{Susskind} But the present calculation shows that
running coupling constant calculations can be consistently carried out
in lowest orders of perturbation theory without inclusion of
perturbative modifications of the vacuum state.  This leads to a new
demand for similar perturbative analysis of theories that may include
effects usually thought to be associated with ground state properties,
the closest being a scalar theory with quartic interaction term and 
infinitesimally small coupling constant.

Obviously, a more sophisticated treatment is necessary in gauge theories
for many reasons but, in particular, because they require an additional
cutoff limiting the longitudinal momentum fractions from below.  The
small-$x$ cutoff appears in the unitary transformation ${\cal
U}_\lambda$ as well as in the transformations ${\cal U}_{\lambda_1}
{\cal U}^\dagger_{\lambda_2}$, the former depending on and the latter
being independent of the ultraviolet cutoff $\Delta$.  This difference
between the small-$x$ cutoff and the ultraviolet cutoff cannot be
studied in the case of scalar fields discussed in the present paper.
However, the third order boost-invariant similarity factors we derived
in the case of $\phi^3$ directly apply, for example, in calculation of a
triple-gluon vertex in effective QCD.

Finally, we wish to stress the difference between the regularized
initial Hamiltonian for bare particles and the small width Hamiltonian
for effective particles, which contains similarity form factors
$f_\lambda$.  The form factors dampen interactions changing invariant
masses by more than $\lambda$ and thus can tame the spread of eigenstate
wave functions for low lying eigenvalues into regions of high relative
momenta of constituents.  This feature may lead to exponential
convergence of the eigenstate expansion in the effective particle basis.
Such convergence is not expected in the case of bare particles.  The
fine structure of effective particles would then unfold in the
transformation ${\cal U}_{\lambda_1} {\cal U}^\dagger_{\lambda_2}$
relating effective degrees of freedom at two different scales, one
corresponding to the binding scale and the other to the high momentum
transfer probe in question.  This picture encourages opinion that 
the present calculation provides a pattern worth trying in
application to realistic theories.

\vskip.3in
{\bf Acknowledgments}
\nopagebreak
\vskip.2in
\nopagebreak

The author thanks Ken Wilson for important discussion concerning 
$x$-dependent counterterms and the manuscript, and Bob Perry for 
his comments.

\vskip.3in
{\bf Appendix}
\nopagebreak
\vskip.2in
\nopagebreak

The loop integrals in Eq.  (2.25) are given in the order implied by
first three terms in the first bracket.  The first term contains the
integral $I_1$, the second term in the bracket together with the first
mass counterterm term contain $I_2$ and the third term in the bracket
together with the second mass counterterm contain $I_3$.  The right-hand
side of Eq.  (2.25) is given by Eq.  (3.6) where $\gamma_\lambda(k_1,
k_2, k_3) = I_1 + I_2 + I_3$. In all three terms we have

$$ {\cal M}^2_{12} = (\kappa^{\perp \, 2}_{12} + \mu^2)/(x_1 x_2) \quad
, \eqno(A.001) $$

\noindent where

$$ x_1 = k_1^+/k_3^+ \quad , \eqno(A.002) $$

\noindent and

$$ \kappa^\perp_{12} = k^\perp_1 - x_1 k^\perp_3 \quad . \eqno(A.003) $$

The first term integral is

$$ I_1 = \left({g \over 2}\right)^2 {1\over 2 (2\pi)^5} {1\over 2}
\left[ \int_{x_1}^1 {dx \over x(1-x)(x-x_1)} \int d^4 \kappa^\perp
{1\over k^{+\,2}_3}\, 8 \, {\cal F}_{3\lambda} (a,b,c,d) \,
r_{\Delta 1} \quad + \quad (1 \leftrightarrow 2) \right] \quad ,
\eqno(A.101) $$

\noindent where in ${\cal F}_{3\lambda}(a,b,c,d)$ in Eqs.
(2.26a-b) one substitutes

$$ ab = - ba = \mu^2 - {\cal M}^2_{68} \quad , \eqno(A.102)$$

$$ ac = - ca = {\cal M}^2_{12} - {\cal M}^2 \quad , \eqno(A.103)$$

$$ ad = - da = {\cal M}^2_{12} - \mu^2 \quad , \eqno(A.104)$$

$$ bc = - cb = {\cal M}^2_{16} - \mu^2 \quad , \eqno(A.105)$$

$$ bd = - db = bc/x + cd = ba/x_2 + ad \quad , \eqno(A.106)$$

$$ cd = - dc = {\cal M}^2 - \mu^2 \quad , \eqno(A.107)$$

\noindent with

$$ {\cal M}^2 = (\kappa^{\perp \, 2} + \mu^2)/[x(1-x)] \quad ,
\eqno(A.108) $$

$$ {\cal M}^2_{16} = x^2(\kappa^{\perp \, 2}_{16} + \mu^2)/[(x-x_1)x_1]
\quad ,\eqno(A.109) $$

$$ {\cal M}^2_{68} = x^2_2(\kappa^{\perp \, 2}_{68} +
\mu^2)/[(x-x_1)(1-x)] \quad , \eqno(A.110)$$

$$ r_{\Delta 1} = \exp{ \left\{- \left[ [\eta(x) + \eta(1-x)]
\kappa^{\perp \, 2} + [\eta[(x-x_1)/x] + \eta(x_1/x)] \kappa^{\perp \,
2}_{16} + \right. \right. }$$
$$ { \left. \left. 
+ [\eta[(x-x_1)/x_2] + \eta[(1-x)/x_2]]\kappa^{\perp \, 2}_{68}
\right] / \Delta^2 \right\} } \quad , \eqno(A.111)$$

\noindent and

$$ \kappa^\perp_{16} = \kappa^\perp_{12} - x_1 \kappa^\perp /x
\quad , \eqno(A.112)$$

$$ \kappa^\perp_{68} = \kappa^\perp - (1-x) \kappa^\perp_{12} /x_2
\quad , \eqno(A.113)$$

$$ P^+_{ab} = P^+_{ba} = x_2 k^+_3 \quad , \eqno(A.114)$$

$$ P^+_{bc} = P^+_{cb} = x k^+_3   \quad , \eqno(A.115)$$

$$ P^+_{bd} = P^+_{db} = P^+_{ca} = P^+_{ac} = P^+_{cd} = P^+_{dc} =
k^+_3 \quad . \eqno(A.116)$$

The expression for $I_2$ is

\begin{eqnarray*}
&& I_2 =
{1\over 2} \left[
\left({g \over 2}\right)^2 {1\over 2 (2\pi)^5}
\int_0^1 {dx \over x(1-x)} \int d^4 \kappa^\perp
\left[ {1\over k^{+\,2}_2}\, 4 \, {\cal F}_{3\lambda}
(a,b,c,d) \right. \right. \\
&& \\
&& \left. \left. - 2 {f_{cd} - 1 \over dc} {1\over x_2} \left[- {2\over
k_2^+}
{\cal F}_{2\lambda_0}(a, b, c)\right]
\right] \, r_{\Delta 2}  - 2 {f_{ad} - 1 \over da} {g \over 2} {\delta
\mu^2_0 \over x_2}
\quad + \quad (1 \leftrightarrow 2) \right] \quad ,
\end{eqnarray*}
$$\eqno(A.201) $$

\noindent where in ${\cal F}_{2\lambda_0}(a, b, c)$ in
Eq.  (2.22) and ${\cal F}_{3\lambda}(a,b,c,d)$ in Eqs.
(2.26a-b) one substitutes

$$ ab = - ba = \mu^2 - {\cal M}^2 \quad , \eqno(A.202)$$

$$ ac = - ca = 0 \quad , \eqno(A.203)$$

$$ ad = - da = cd \quad , \eqno(A.204)$$

$$ bc = - cb = ba \quad , \eqno(A.205)$$

$$ bd = - db = bc/x_2 + cd \quad , \eqno(A.206)$$

$$ cd = - dc = {\cal M}^2_{12} - \mu^2 \quad , \eqno(A.207)$$

\noindent with the same

$$ {\cal M}^2 = (\kappa^{\perp \, 2} + \mu^2)/[x(1-x)] \quad ,
\eqno(A.208) $$

\noindent and

$$ r_{\Delta 2} = \exp{ \left\{ - 2 [\eta(x) + \eta(1-x)] \kappa^{\perp
\, 2} / \Delta^2 \right\} } \quad , \eqno(A.209)$$

$$ P^+_{ab} = P^+_{ac} = P^+_{bc} = x_2 k^+_3 \quad , \eqno(A.210)$$

$$ P^+_{ad} = P^+_{bd} = P^+_{cd} = k^+_3 \quad . \eqno(A.211)$$

The expression for $I_3$ is

\begin{eqnarray*}
&& I_3 = \left({g \over 2}\right)^2 {1\over 2 (2\pi)^5}
\int_0^1 {dx \over x(1-x)} \int d^4 \kappa^\perp
\left[ {1\over k^{+\,2}_3}\, 2 \, {\cal F}_{3\lambda}
(a,b,c,d) \right. \\
&& \\
&& \left. + {f_{ad} - 1 \over da} \left[- {2\over
k_3^+}
{\cal F}_{2\lambda_0} (b, c, d)\right]
\right] \, r_{\Delta 3}  + {f_{ad} - 1 \over da} {g \over 2} {\delta
\mu^2_0 }
 \quad ,
\end{eqnarray*}
$$\eqno(A.301) $$

\noindent where in ${\cal F}_{2\lambda_0}
(b, c, d)$ in
Eq.  (2.22) and ${\cal F}_{3\lambda}(a,b,c,d)$ in Eqs.
(2.26a-b) one substitutes

$$ ab = ad = {\cal
M}^2_{12} - \mu^2  \quad , \eqno(A.302)$$

$$ ac = {\cal M}^2_{12} - {\cal M}^2  \quad , \eqno(A.303)$$

$$ bc =  dc = \mu^2 - {\cal M}^2 \quad , \eqno(A.304)$$

$$ bd = 0 , \eqno(A.305)$$

\noindent with the same $ {\cal M}^2 $ as for $I_1$ and $I_2$ and

$$ r_{\Delta 3} = r_{\Delta 2} \quad , \eqno(A.306)$$

\noindent and all parent momenta equal $k_3$.

\end{document}